\documentclass[10pt]{article}
\usepackage{graphicx}
\usepackage{amssymb,amsmath}
\usepackage{float}
\usepackage{tcolorbox}
\usepackage{amsthm}
\usepackage{graphicx}
\usepackage{hyperref}
\hypersetup{
	colorlinks=true,
	linkcolor=blue,
	filecolor=magenta,      
	urlcolor=cyan}

\begin{document}  
\title{Thomson problem in the disk }
\author{
  Paolo Amore \\
  \small Facultad de Ciencias, CUICBAS, Universidad de Colima,\\
  \small Bernal D\'{i}az del Castillo 340, Colima, Colima, Mexico \\
  \texttt{paolo@ucol.mx}
  \and
  Ulises Zarate \\
  \small Facultad de Ciencias, Universidad de Colima,\\
  \small Bernal D\'{i}az del Castillo 340, Colima, Colima, Mexico \\
  \texttt{uzarate@ucol.mx}
}

\maketitle

\begin{abstract}
We investigate the classical ground state of a large number of charges confined inside a disk and interacting via 
the Coulomb potential. By realizing the important role that the peripheral charges play in determining the lowest energy solutions, we have successfully implemented an algorithm that allows us to work with configurations with a desired number of border charges. 
This feature brings a consistent reduction in the computational complexity of the problem, thus simplifying the search of global minima of the energy. Additionally, we have implemented a {\sl divide and conquer} approach  which has allowed us to study configurations of size never reached before (the largest one corresponding to $N=40886$ charges). These last configurations, in particular, are seen to display an increasingly rich structure of topological defects as $N$ gets larger.
\end{abstract}

\maketitle

\section{Introduction}
\label{sec:intro}

In this paper we study the configurations of a large number of classical Coulomb charges confined  to the interior of a disk (without loss of generality we  assume unit radius). There has been considerable interest in the past in studying this system (see for instance refs.~\cite{Bedanov94,Bedanov95,Oymak01,Worley06,Moore07,Olvera13,Cerkaski15,Cerkaski17a,Cerkaski17b,Amore17}), particularly in  relation to the onset of {\sl Wigner crystallization}.
The equilibrium configurations for this system are non--uniform and qualitatively well described by a conformal crystal, with disclinations and dislocations playing an important role in determining configurations of minimal energy~\cite{Moore07}. The regime $N \gg 1$, in particular, is interesting because it may allow to compare the low energy configurations solutions to the discrete problem with approximations based on the continuum~\cite{Grason21}.

This problem can also be considered a variant of the more famous Thomson problem, which concerns the equilibrium configurations
of a number of charges on the surface of a sphere~\cite{Erber91,Glasser92,Bergersen94,Erber97,Saff94,Saff94b,Saff97,Altschuler97,Bowick02,Wales06,Wales09}.
The charge density for the two systems, the disk or the sphere,  however, has a very different behavior: in the former, the continuum charge distribution is non--uniform and larger at the border, whereas in the latter the continuum distribution is uniform. In both cases however the impossibility to cover the whole region with hexagonal cells (as it would happen on the infinite plane) leads to the appearance of a fixed topological charge ($12$ for the sphere and $6$ for the disk), a direct consequence of Euler's theorem of topology. This charge is induced by the curvature for the case of the sphere and by the geometrical frustration caused by the border for the case of the disk. While the net topological charge in each domain is constant, the structure of defects becomes increasingly rich as the density is increased (see in particular ref.~\cite{Wales06,Wales09} for the case of the sphere).

Both examples, Thomson's problem inside the disk or on the sphere, can be regarded as special cases of two dimensional matter, i.e. systems of particles interacting with a given potential (not necessarily Coulombian) on a frozen topology. In this case, the  topology of the surface  induces peculiar features in the ground state configurations of the system~\cite{Bowick02,Bowick00,Bowick03,Bowick06,Bowick07,Bowick09,Rad11,Wales13}.

The main goal of our work is to present efficient algorithms for studying large configurations of Coulomb charges inside a disk and then compare the numerical results obtained in this way with the best results available in the literature. In addition to this, the algorithms that we have devised also allow us to extend our study to consider much denser configurations, never before considered. 

The paper is organized as follows: in section \ref{sec:algo} we describe the numerical algorithms that we have devised, in section \ref{sec:results} we present the numerical results that we have obtained; finally in section \ref{sec:concl} we  draw our conclusions.

\section{Numerical algorithm}
\label{sec:algo}

We consider $N$ charges inside a disk (for convenience we set the radius of the disk to $1$)
interacting via the Coulomb potential; the total energy of this system is 
\begin{equation}
E = \sum_{i=2}^N \sum_{j=1}^{i-1} \frac{1}{r_{ij}}
\label{energy}
\end{equation}
with $\vec{r}_i \equiv (x_i,y_i)$ and $r_{ij} =|\vec{r}_i - \vec{r}_j|$.

It is well known that finding the global minimum of eq.~(\ref{energy}) is prohibitively difficult for large values of $N$. This feature is shared with problems of similar nature, such as the Thomson problem on the sphere, which are known to be NP-hard~\cite{Wille85A,Wille85B}. In both cases, the number of local minima is observed to grow exponentially with $N$, making the full exploration of the energy landscape associated with the problem a very difficult task.

While one expects the problems to be of comparable difficulty and share some similar feature, the minimization of eq.~(\ref{energy}) can be greatly facilitated by devising algorithms that take advantage of specific properties.

The continuum limit of the solutions that minimize eq.~(\ref{energy}), for instance, is a non--homogeneous configuration with a density that is larger at the border~\cite{Moore07} (conversely, for the Thomson problem, the charges tend to distribute uniformly as $N \rightarrow \infty$). For finite $N$ the global minimum of eq.~(\ref{energy}) corresponds to configurations with a number of border charges that is proportional to $N^{2/3}$~\cite{Worley06,Moore07}. 

The efficiency of any numerical approach will be limited, if the algorithm cannot produce easily configurations with the correct number of border charges. One example of this occurrence is found in ref.~\cite{Cerkaski15} where the minimum for $N=395$ charges was not correctly identified numerically~\cite{Amore17}~\footnote{Ref.~\cite{Cerkaski17a}, the reply to ref.~\cite{Amore17}, also misidentified the energy of the ground state for $381$ charges to be $E=102764.53$. We have found a configuration with  a lower energy, $E=102764.3757$. The reader may find this configuration in the supplemental material of our paper~\cite{AZ_suppA,AZ_suppB}. These examples help to understand the difficulty in finding the lowest energy configurations, even for not so large $N$.}. 

In general an algorithm will tend to produce a configuration with a given number of border charges $N_b$ with some probability: in our experience the probability distribution is almost a Gaussian with the position of the peak showing sensitivity to the initial random distribution being used. In particular, if the initial random configuration is not selected appropriately the probability of producing a final configuration with the proper $N_b$ could be {\sl exponentially} suppressed. In this case one needs a exponentially large number of trials to get to the global minimum. 
This exponential increase in difficulty however can be avoided if the algorithm is capable from the start of producing {\sl only} configurations with 
the desired $N_b$, as done  by one of us in ref.~\cite{Amore17}. To avoid misunderstandings, fixing the proper $N_b$  makes the search exponentially more effective but it does not make the problem trivial since the subset of solutions with a given $N_b$ also grows exponentially large as $N$ grows.

\begin{figure}
\begin{center}
\bigskip\bigskip\bigskip
\includegraphics[width=8cm]{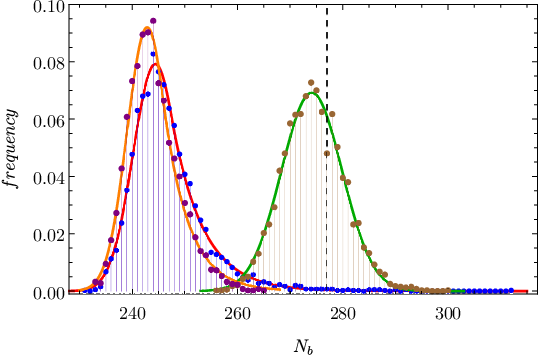}\hspace{1cm}
\bigskip
\caption{Histograms showing the frequency at which configurations with given $N_b$ are obtained starting from a random configuration for $N=1000$. The points correspond to using random initial configurations where the angles in the parametrization are chosen uniformly random on $0 \leq t \leq \pi/2$ and $0 \leq u \leq 2\pi$ (violet points), the points are distributed uniformly on the domain (blue points) and the points are distributed according to the classical charge density (brown points). The vertical dashed line  is the expected number of border charges for the global minimum.}
\label{Fig_Histo_nb}
\end{center}
\end{figure}

In Fig.~\ref{Fig_Histo_nb} we show the histograms for  the frequency at which configurations with given $N_b$ are obtained starting from a random configuration for $N=1000$.  As one can see from this figure the choice of the initial random distribution greatly affects the efficiency in 
producing configurations with the appropriate number of border charges: only for the case of points distributed according to the continuum distribution for this problem the peak of the Gaussian is very close to the optimal $N_b$, whereas in the other two cases the probability of producing the correct $N_b$ is very small.

With the previous experience accumulated in ref.~\cite{Amore17} we have devised an algorithm that allows one to target configurations with a given number of border charges. In what follows we  discuss the algorithm  in some detail.

As a starting point, we must recognize that we are dealing with a constrained minimization problem, because the charges are confined to a disk. 
A point in the disk can however be represented in polar--like coordinates as
\begin{equation}
(x,y) = \sin^2 t \ (\cos u, \sin u )
\label{parametrization}
\end{equation}
where $0 \leq t  \leq \pi/2$ and $0 \leq u \leq 2 \pi$. By using this parametrization we manage to transform our problem into an unconstrained one, which is easier to handle. Notice that eq.~(\ref{energy}) is now a function of $2N$ angles.

Assuming that we want $N_b$ charges on the border, we may use the parametrization of (\ref{parametrization}) in two different forms:
\begin{equation}
\begin{split}
(x,y) &= \sigma \sin^2 t \ (\cos u, \sin u ) \hspace{1cm} , \hspace{1cm} {\rm internal \ \ points} \\
(x,y) &= (\cos u, \sin u ) \hspace{2.1cm} , \hspace{1cm} {\rm border \ \ points}
\end{split}
\label{parametrization2}
\end{equation}
where $0 < \sigma < 1$ is a parameter that constrains the internal points to the interior of the disk. The value of $\sigma$ should be determined empirically and  in our experience $\sigma  \approx  1 - \frac{1}{2\sqrt{N}}$ works very well~\footnote{Choosing $\sigma$ too small or too large will produce "artificial" equilibrium configurations, i.e. configurations that are not at equilibrium when $\sigma$ is eventually set to $1$.}.
A supplemental advantage of this parametrization consists of reducing the total number of degrees of freedom (dof) from $2N$ to $2 N-N_b$.

Once the total energy of the system is expressed in terms of eq.~(\ref{parametrization2}) one can look for its minima using standard procedures. In our case, we have implemented the minimization following two different procedures: applying the truncated Newton method (TN) to initial random configurations with fixed $N_b$ and using the basin--hopping (BH) method~\cite{Li87,Wales97,Wales99,Wales03}, which is widely considered the most effective algorithm for the search of global minima. As we will see in the next section, where we discuss the numerical results, both algorithms turn out to be very effective in finding good candidates for the global minimum of the energy.

One of the goals of the present paper, however, is to try to study configurations of much larger size than the ones found in the literature: to the best of our knowledge, $N=5000$ is the largest configuration that has been previously studied, in ref.~\cite{Moore07}, although unfortunately the value of its energy is not reported. For such large $N$ not only the minimization process can be very time--consuming but additionally round--off errors may become increasingly large, reducing the effectiveness of our approach.

\begin{figure}
\begin{center}
\bigskip\bigskip\bigskip
\includegraphics[width=8cm]{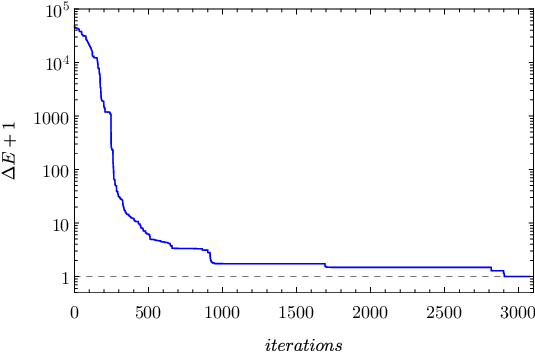}
\bigskip
\caption{Energy of the configuration of $1000$ charges obtained with a single run of the D\&C algorithm (corresponding in this case to about $3100$ iteration).  $\Delta E$ is the difference between the energy at a given iteration and the final energy obtained. The vertical axis plots $\Delta E  + 1$ to avoid the last point running to $-\infty$.}
\label{Fig_energy_dc_1000}
\end{center}
\end{figure}

For these reasons we have implemented  a {\sl divide and conquer} (D\&C) approach, where again the minimization can be performed using TN or BH.
It is worth describing our approach in some detail.

The approach consists of the following steps:
\begin{itemize}
\item instead of performing the minimization over the whole disk, one can select a smaller portion of it, such a disk of smaller radius, centered at a random point inside the unit disk; we have found that it is convenient to use the classical charge distribution inside the disk as the probability distribution for such points.

\item the original $2N-N_b$ dofs can then be partitioned into two sets of points, depending whether the points fall outside or inside the smaller disk;
in this case $2N-N_b= N_{out} +N_{in}$, where $N_{out}$ and $N_{in}$ are the numbers of external and internal dofs.
\item the external points are fixed at their positions, while the internal points are free to move and may be initially perturbed to allow generate a configuration out of equilibrium; because the equilibrium distribution of points inside the disks is not homogeneous (as it would be for the Thomson problem on the sphere), the amount of perturbation depends on the 
position of the smaller disk (if this disk is close to the border of the region, the perturbation should be smaller to take into account the fact that charges are closer); the energy functional is then splitted into three contributions, $E = E_{in-in} +  E_{out-out} +E_{in-out} $ corresponding to the  electrostatic energy of the internal charges ($E_{in-in}$),  of the external charges ($E_{out-out}$) and to the interaction energy between internal and external charges; typically we want $N_{in} \ll N_{out}$ and as a result we expect $E_{in-in} \ll E_{in-out} \ll E_{out-out}$. Notice that $E_{out-out}$ needs to be calculated only once at this stage, since it depends on the dofs which are kept fixed;
\item while running the algorithm the radius of the smaller disk can be progressively increased from some smaller value (typically $r =0.1$) to a larger one (typically $r \approx 1$); these values are not sacred but they should be changed keeping in mind that as one starts, it is possible to find good improvements to the energy while working with very few degrees of freedom; as the algorithm advances, a larger number of dofs should be used to allow the change in energy to be safely above round--off errors and thus provide sizable improvement to the total energy;
\item upon minimization a new configuration is found where the internal points are in equilibrium (but this will not necessarily correspond to an equilibrium configuration of the whole system once all the dofs are unfrozen): if the energy $E_{in-in}+E_{in-out} $ has a lower value the configuration is retained otherwise it is discarded;
\item the new configuration thus obtained  is compared with the immediately previous  configuration, identifying the charge that has been most displaced in the minimization process: if the displacement is above a threshold established by the user, the center of the smaller disk is placed at this point and the minimization process is repeated; we call this step {\sl defect chasing}, because  sufficiently large changes in the positions of the charges induce a change in the local structure of the Voronoi cells (while leaving the total topological charge unchanged); if the minimization in the new region is again successful, the {\sl defect chasing} can be repeated, until no better minimum is found, or the displacement threshold is not met;

\item repeat the process starting from the first step a large number of times until the changes in the energy are sufficiently small; the algorithm can also be stopped at any time, if needed, and in that case a minimization on the full domain should still be performed to make sure that the gradient is small enough.
\end{itemize}

In \cite{AZ_suppAnim} the reader can find an animation corresponding to the application of the D\&C algorithm to a configuration with $2000$ charges inside the disk. The red points in the animation are the active dofs in the minimization process.

In Fig.~\ref{Fig_energy_dc_1000} we show an example of application of this algorithm for the case of $1000$ charges starting from a completely random initial configuration.  The final configuration is the actual best minimum that we have found for this case.
One of the advantages of the D\&C algorithm is the ability to better explore the landscape without being easily trapped by a local minimum, as it would happen if the minimization would be carried out on the whole domain.

As we will comment later, {\sl all} the records that we have obtained for the configurations with $N \geq 1000$ 
correspond to using the D\&C algorithm, although we have also used intensively the basin--hopping algorithm.
In our experience D\&C outperforms more standard approach for very large configurations of Coulomb charges inside a disk~\footnote{Incidentally, motivated by our success we have also applied it to the standard Thomson problem on the sphere, but in this case the effectiveness of the D\&C appears to be smaller, possibly due to the absence of a border. Further experimentation for this case is needed.}.

\section{Numerical results}
\label{sec:results}

We have used the algorithms described  in the previous section to calculate a large number of configurations: our data include 
all configurations with $N \leq 330$  and selected ones with larger $N$. The largest configuration that we have studied is  $N=40886$, which is significantly larger that any other one previously reported in the literature~\footnote{To the best of our knowledge the largest configuration previously studied corresponds to $N=5000$, which was considered by Mughal and Moore~\cite{Moore07}. Unfortunately they did not report the value of the energy they found, so that a direct comparison of our result for $N=5000$ with theirs is not immediate. An indirect comparison can however be made by noticing that the fit reported in \cite{Moore07} provides an energy about $82.2$ units above our best value. }.

In general our calculations either reproduce or improve all results in the literature for which numerical values are available, but also  extend to much larger $N$ than previously studied. 

Understanding the particular features of these large configurations may be a useful in assessing the effectiveness of the models based on the continuum.

Worley~\cite{Worley06} and Mughal and Moore~\cite{Moore07} have found that the number of border charges scales as $N_b \propto N^{2/3}$ for $N \gg 1$.
The argument to understand this behavior is straightforward: assume that the charge is distributed according the continuum density:
\begin{equation}
\rho = \frac{N}{2\pi} \frac{1}{\sqrt{1-r^2}} \ ,
\label{eq:rho}
\end{equation}
and calculate $N_b$ as the charge contained in a circular annulus of width $\delta r$ and outer radius $1$:
\begin{equation}
N_b =  N \sqrt{1- (\delta r)^2} \ .
\label{eq:Nb}
\end{equation}

Assuming that the border charges are evenly distributed on the border we estimate the interparticle distance on the border to be
$d=2\pi/N_b$; it is reasonable to expect that $\delta r = \frac{2\pi\eta}{N_b}$, with $\eta>0$. 
Upon substitution of this relation inside (\ref{eq:Nb}) and assuming $N \gg 1$ one obtains the relation
\begin{equation}
N_b \approx (4 \pi  \eta  N^2)^{1/3} \ .   
\end{equation}

We have fitted the observed values of $N_b$ for the numerical solutions with $N \leq 5000$ with the form
\begin{equation}
N_b^{\rm fit}(N) \approx   2.84328 N^{2/3} - 0.530196 N^{1/3}  -2.32866  \ .
\label{eq:nbfit} 
\end{equation}

\begin{figure}
\begin{center}
\bigskip\bigskip\bigskip
\includegraphics[width=8cm]{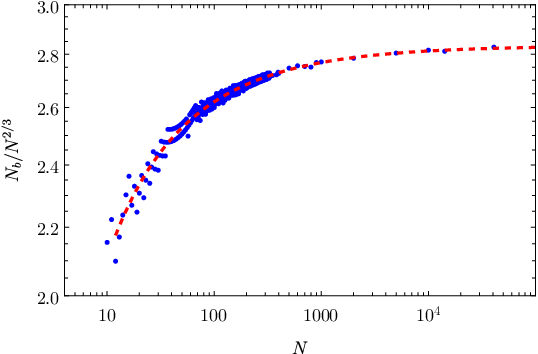}
\bigskip
\caption{$N_b/N^{2/3}$ as a function of $N$. The blue points are the values obtained from the numerical calculation, the dashed red curve corresponds to the fit. }
\label{Fig_nb}
\end{center}
\end{figure}

\begin{figure}
\begin{center}
\bigskip\bigskip\bigskip
\includegraphics[width=8cm]{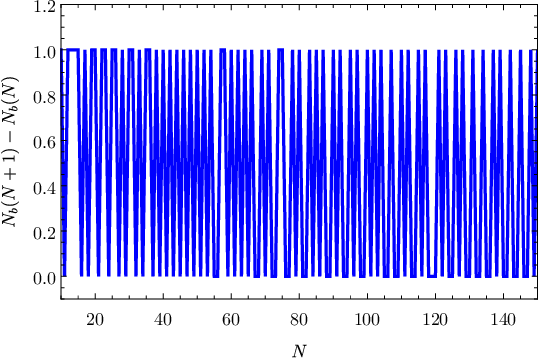}
\bigskip
\caption{$N_b(N+1)-N_b(N)$ as a function of $N$. }
\label{Fig_deltanb}
\end{center}
\end{figure}

In Fig.~\ref{Fig_nb} we display $N_b/N^{2/3}$ as a function of $N$: here the points represent the values obtained with the numerical calculation, while the dashed curve represent the fit of eq.~(\ref{eq:nbfit}). 
In Fig.~\ref{Fig_deltanb} we plot $N_b(N+1)-N_b(N)$ as function of $N$. This quantity normally alternates between the values $0$ and $1$, with sporadic repetitions which correspond to maintaining the same $N_b$ for three successive configurations, or, alternatively, increasing the $N_b$ of one unit for three successive configurations (the latter phenomenon however occurs only for rather small, $N =74$ being the last $N$ where we have observed it).

For large $N$ the fit of eq.~(\ref{eq:nbfit}) can help to restrict the search for optimal configuration to a rather small region of values of $N_b$, thus allowing one to considerably speed up the numerical calculation.

The energy of a configuration of $N$ charges inside a unit disk mutually repelling through the Coulomb potential for $N \rightarrow \infty$ is expected to behave as~\cite{Oymak01,Worley06, Moore07}
\begin{equation}
E^{(fit)}(N) \approx \kappa_1 N^2 + \kappa_2 N^{3/2} + \kappa_3 N + \kappa_4 \sqrt{N} + \kappa_5 + \dots
\label{eq:enmughal}
\end{equation}
where $\kappa_1 = \pi/4$. Mughal and Moore have estimated the remaining coefficients in ref.~\cite{Moore07} using their numerical results.

The fit of our data from $N=100$ to $N=5000$ gives the coefficients
\begin{equation}
\begin{split}
\kappa_2 &= -1.5628 \\
\kappa_3 &=  1.0302 \\
\kappa_4 &= -0.9899 \\
\kappa_5 &=  4.9255 \\
\end{split}
\label{eq_fit}
\end{equation}
which can be compared with those reported in ref.~\cite{Moore07}. In particular the values for $\kappa_2$ and $\kappa_3$ that we have obtained are close to those of Mughal and Moore, whereas $\kappa_4$ and $\kappa_5$ are rather different.  We believe that in order to reach a reliable determination of $\kappa_4$ and $\kappa_5$ one should  have at one's disposal a larger set of numerical results than the one we have. Eq.~(\ref{eq:enmughal}) should be intended to provide an approximate estimate of the energy of the configurations at different $N$.

\begin{figure}
\begin{center}
\bigskip\bigskip\bigskip
\includegraphics[width=8cm]{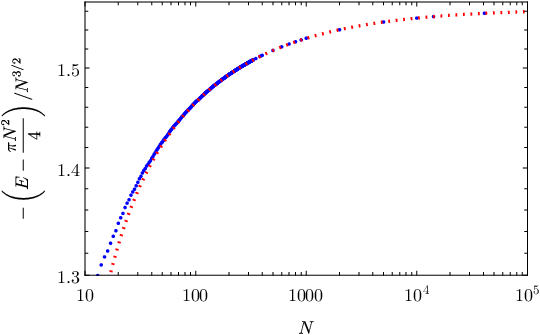}\hspace{1cm}
\bigskip
\caption{$- \left( E(N) -\frac{\pi}{4} N^2\right)$ as a function of $N$. The blue points are the numerical results, while the red curve is the fit of eq.~(\ref{eq:enmughal}) with the coefficients given in eq.~(\ref{eq_fit}).}
\label{Fig_en_asym}
\end{center}
\end{figure}

One expects that the charge  distribution in the discrete model approaches the continuum limit of eq.~(\ref{eq:rho}) for sufficiently large $N$.
The charge density in the discrete model can be obtained directly in terms of the Voronoi tessellation~\cite{Grason21}:
\begin{equation}
\rho^{(discrete)}(\vec{r}_i) = \frac{1}{A_i} \ ,
\end{equation}
where $A_i$ is the area of the Voronoi cell corresponding to the $i^{th}$ charge.

We have found that at finite $N$ the density can be described very well by:
\begin{equation} 
\rho(r) = \rho_{int}(r) + \rho_{b}(r) \ ,
\end{equation}
where $\rho_{int}(r)$ and $\rho_{b}(r)$ are the densities of internal and border charges respectively, given by the expressions
\begin{equation} 
\begin{split}
\rho_{int}(r) &= \frac{N-N_b}{2\pi} \frac{1}{\sqrt{1-r^2}} \left[ \alpha -(\alpha -1) (\beta +1) \left(1-r^2\right)^{\beta/2}
\right] \\
\rho_{b}(r) &= \frac{N_b}{2\pi} \delta(1-r)
\end{split} \ ,
\label{eq:rhofit}
\end{equation}
where
\begin{equation}
\begin{split}
\int_{disk} \rho_{int}(r) d^2r &= (N-N_b) \\
\int_{disk} \rho_{b}(r) d^2r &= N_b \\
\end{split} \ .
\end{equation}

The parameters $\alpha$ and $\beta$ should be adjusted at each $N$ to provide the best fit of the numerical data and they should 
be such that the continuum limit (\ref{eq:rho}) is recovered for $N \rightarrow \infty$.  
By noticing that $\lim_{N \rightarrow \infty} \frac{N_b}{N} = 0$, we  see that  the density reduces to eq.~(\ref{eq:rho}) at large 
$N$ if $\lim_{N \rightarrow \infty} \alpha=1$.

\begin{figure}
\begin{center}
\bigskip\bigskip\bigskip
\includegraphics[width=8cm]{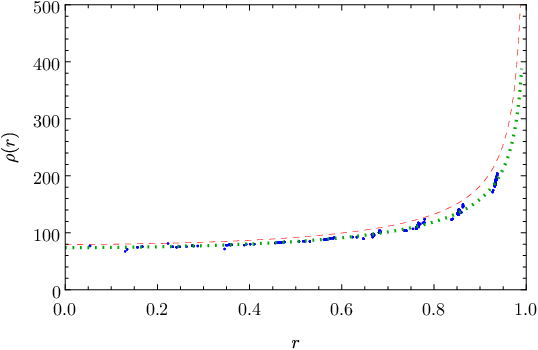}\hspace{1cm}
\bigskip
\caption{Charge density for the optimal configuration with $500$ point charges in the unit circle. The blue points are the 
values obtained from the numerical calculation, the dashed red curve corresponds to eq.~(\ref{eq:rho}); the dotted green line is the fit
(\ref{eq:rhofit}) for $\rho_{int}(r)$: $\rho_{int}(r) = \frac{327}{2 \pi  \sqrt{1-r^2}} \left(1.5147\, -\frac{0.0932257}{\left(1-r^2\right)^{0.409437}}\right)$.}
\label{Fig_rho500}
\end{center}
\end{figure}

\begin{figure}
\begin{center}
\bigskip\bigskip\bigskip
\includegraphics[width=8cm]{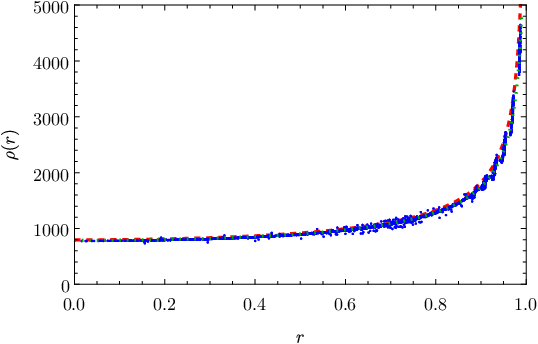}\hspace{1cm}
\bigskip
\caption{Charge density for the optimal configuration with $5000$ point charges in the unit circle. The blue points are the 
	values obtained from the numerical calculation, the dashed red curve corresponds to eq.~(\ref{eq:rho}); the dotted green line is the fit
	(\ref{eq:rhofit}) for $\rho_{int}(r)$: $\rho_{int}(r) = \frac{2090}{\pi  \sqrt{1-r^2}} \left(1.21875\, -\frac{0.0372972}{\left(1-r^2\right)^{0.414748}}\right)$.}
\label{Fig_rho5000}
\end{center}
\end{figure}

In Figs.~\ref{Fig_rho500} and \ref{Fig_rho5000} we plot the charge density for the best configuration that we have found for $N=500$ and $N=5000$.  The dashed red curve corresponds 
to the continuum density of eq.~(\ref{eq:rho}), whereas the dotted green curve is the  fit (\ref{eq:rhofit}) for $\rho_{int}(r)$.

\begin{figure}
\begin{center}
\bigskip\bigskip\bigskip
\includegraphics[width=8cm]{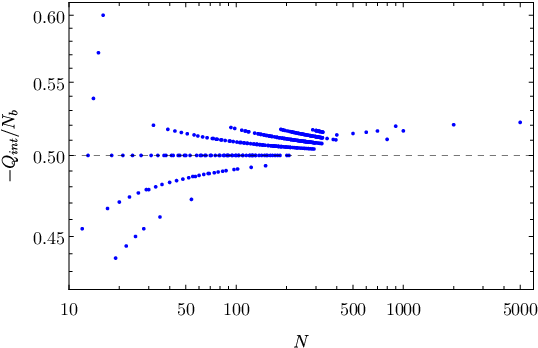}\hspace{1cm}
\bigskip
\caption{$-Q_{int}/N_b$ as a function of the number of charges.}
\label{Fig_QintoverNb}
\end{center}
\end{figure}

In Fig.~\ref{Fig_QintoverNb} we plot  $-Q_{int}/N_b$ as a function of $N$, where $Q_{int}$ is the total internal topological charge in the disk.
From this plot we see that $Q_{int} \approx N_b/2$: the reason for this behavior is the fact that typically the Voronoi cells at the border of the disk are pentagons and quadrilaterals, in an alternating sequence, with possible sporadic repetitions of a pentagon (lower sequences) or a quadrilateral (upper sequences).

\begin{figure}
\begin{center}
\bigskip\bigskip\bigskip
\includegraphics[width=8cm]{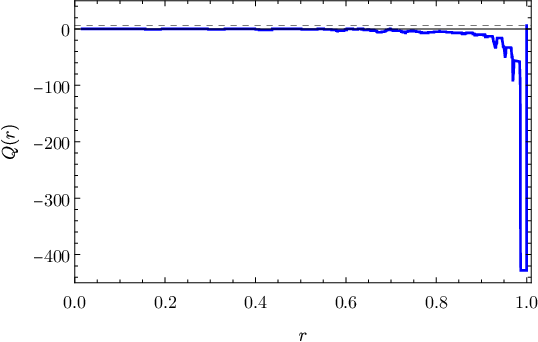}\hspace{1cm}
\bigskip
\caption{Topological charge in a disk of radius $r$ as function of $r$ for $N=5000$.}
\label{Fig_Qtopo}
\end{center}
\end{figure}

In Fig.~\ref{Fig_Qtopo} we plot the topological charge contained from $0$ to $r$, as a function of $r$. For $r=1$, the topological charge is
$Q(1) = Q_{tot} =6$, as required from Euler's theorem (the dashed line corresponds to $Q=6$). The topological charge reaches a minimum very close to the border, after which it rapidly grows finally reaching the final value at the border $Q_{tot}=6$. 
Because the minimum is reached at the very last layer of cells before the border, it actually corresponds to $Q_{int}$ plotted in Fig.~\ref{Fig_Qtopo}.

\begin{figure}
\begin{center}
\bigskip
\includegraphics[width=8cm]{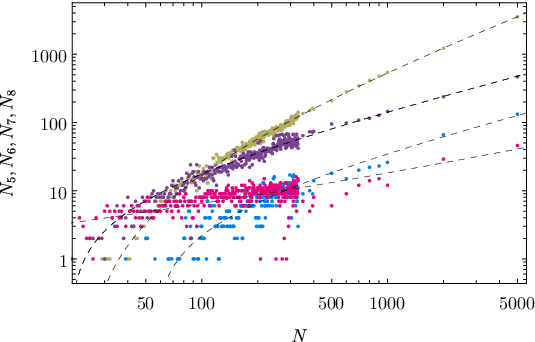}\hspace{1cm}
\bigskip
\caption{Number of pentagonal, hexagonal, heptagonal and octagonal internal Voronoi cells as functions of $N$.}
\label{Fig_N5678}
\end{center}
\end{figure}

\begin{figure}
\begin{center}
\includegraphics[width=6cm]{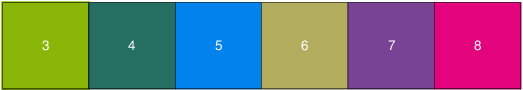}\\
\caption{Color scheme used for the  Voronoi cells: the number within each square represents the number of sides of the Voronoi cell.}
\label{Fig_colorscheme}
\end{center}
\end{figure}

In Fig.~\ref{Fig_N5678} we plot the number of pentagonal, hexagonal, heptagonal and octagonal internal Voronoi cells as functions of $N$, following the color scheme in Fig.~\ref{Fig_colorscheme}. The dashed lines are the fits
\begin{equation}
\begin{split}
N_5^{(fit)}(N) &= 0.555351 N^{2/3}-2.10709 \sqrt[3]{N} \\
N_6^{(fit)}(N) &= N -5.43617 N^{2/3}+7.35492 \sqrt[3]{N} \\
N_7^{(fit)}(N) &= 1.92787 N^{2/3}-5.16971 \sqrt[3]{N} \\
N_8^{(fit)}(N) &= 0.079599 N^{2/3}+1.02801 \sqrt[3]{N} \\
\end{split} \ .
\end{equation}

The particular choice of exponents in the fits is motivated by the fact that $N_5+N_6+N_7+N_8 = N-N_b$ (we have verified that 
 $N^{(fit)}_5+N^{(fit)}_6+N^{(fit)}_7+N^{(fit)}_8 \approx N-N_b$). Additionally with these fits we can obtain the approximate behavior of 
the total internal topological charge:
\begin{equation}
Q_{int} \approx N^{(fit)}_5-N^{(fit)}_7-2 N^{(fit)}_8 = -1.53172 N^{2/3} + 1.0066 \sqrt[3]{N} \ ,
\end{equation}
which reproduces quite well the observed values of $Q_{int}$.

Incidentally, Fig.~\ref{Fig_Qtopo} and the formula above disprove the observation of \cite{Olvera13} that $Q_{int}$ scales linearly with $N$.
The reader should also notice that $N_7 \gg N_5$: this is due to the fact that the last layer of Voronoi cells before the border typically contains a large number of heptagons (pentagons, on the other hand, are present in large numbers on the border, but right now we are only discussing internal cells so they do not contribute to $N_5$).

\begin{figure}
\begin{center}
\bigskip\bigskip\bigskip
\includegraphics[width=5cm]{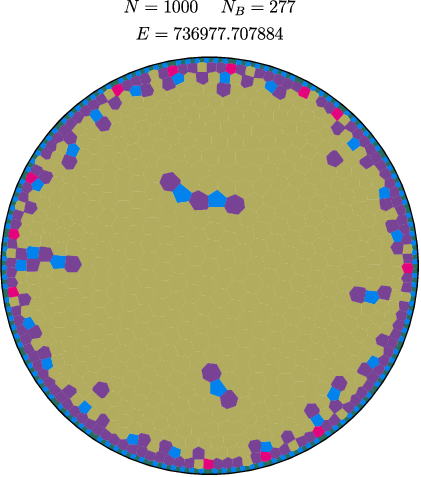} \hspace{1cm}
\includegraphics[width=5cm]{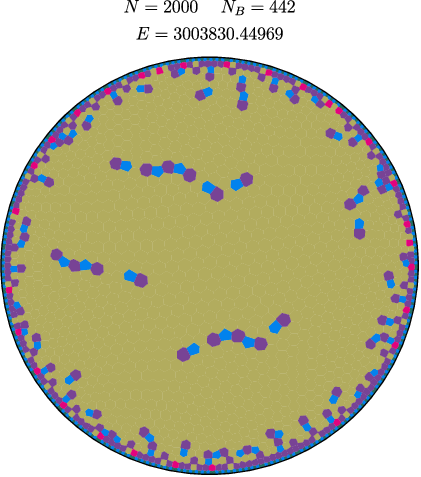}\\
\bigskip
\caption{Best configurations with $1000$ and $2000$ charges found with the D\&C algorithm.}
\label{Fig_conf_1000_2000}
\end{center}
\end{figure}

\begin{figure}
\begin{center}
\bigskip\bigskip\bigskip
\includegraphics[width=5cm]{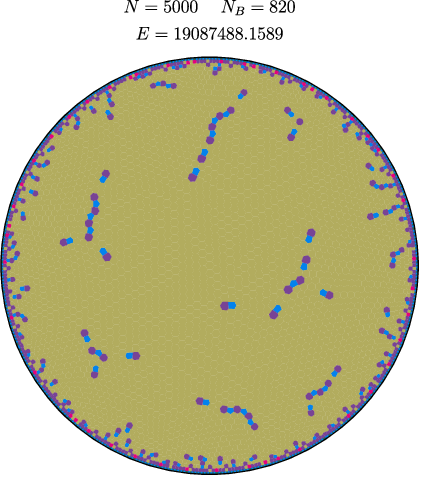} \hspace{1cm}
\includegraphics[width=5cm]{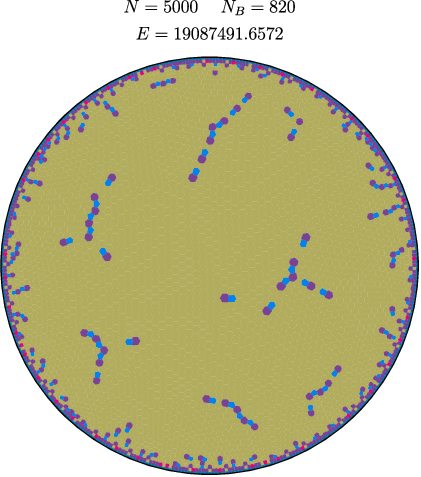}\\
\bigskip
\caption{Best configurations (left plot) and a low energy configuration (right plot) configurations with $5000$ charges found with the D\&C algorithm. }
\label{Fig_conf_5000}
\end{center}
\end{figure}

\begin{figure}
	\begin{center}
		\bigskip\bigskip\bigskip
		\includegraphics[width=5cm]{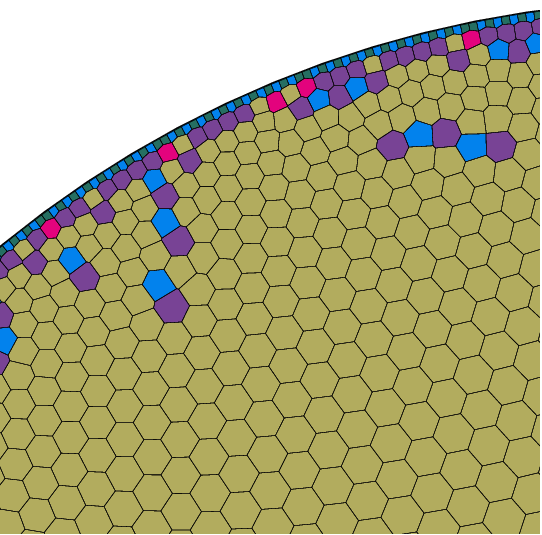} \hspace{1cm}
\includegraphics[width=5cm]{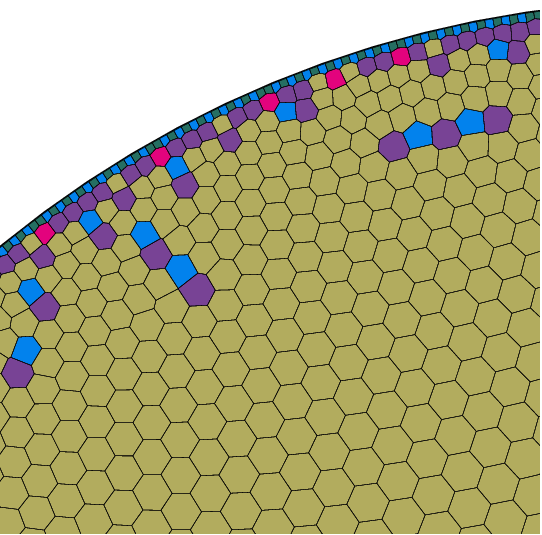}\\
\bigskip
\caption{Detail of the plots in Fig.~\ref{Fig_conf_5000}. }
		\label{Fig_conf_5000_detail}
	\end{center}
\end{figure}

\begin{figure}
	\begin{center}
		\bigskip\bigskip\bigskip
		\includegraphics[width=6cm]{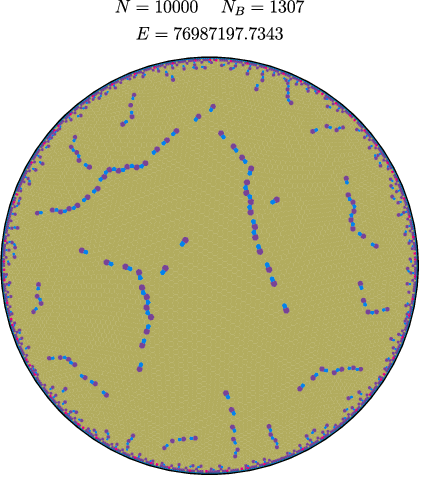}\hspace{1cm}
		\bigskip
		\caption{Best configurations with $10000$ charges found with the D\&C algorithm.}
		\label{Fig_conf_10000}
	\end{center}
\end{figure}

\begin{figure}
\begin{center}
\bigskip\bigskip\bigskip
\includegraphics[width=6cm]{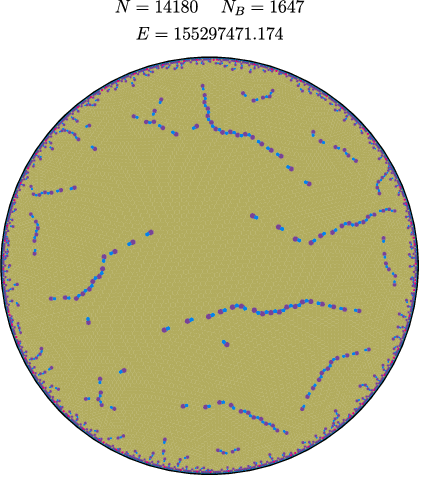}\hspace{1cm}
\bigskip
\caption{Best configurations with $14180$ charges found with the D\&C algorithm.}
\label{Fig_conf_14180}
\end{center}
\end{figure}

\begin{figure}
\begin{center}
\bigskip\bigskip\bigskip
\includegraphics[width=6cm]{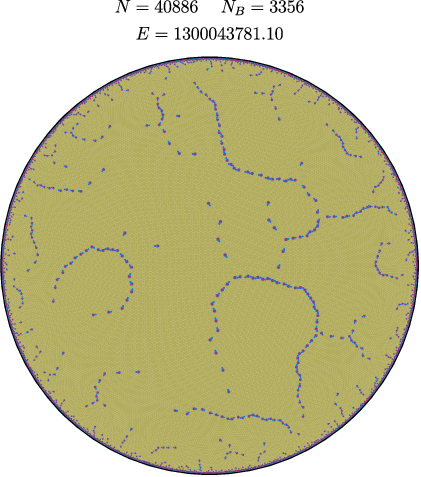}\hspace{1cm}
\bigskip
\caption{Best configurations with $40886$ charges found with the D\&C algorithm.}
\label{Fig_conf_40886}
\end{center}
\end{figure}

Although we have calculated a large number of configurations, it is clearly impossible to show them all here, therefore we will limit ourselves to 
show the largest configurations that we have calculated, $N=1000$ and $2000$ in Fig~\ref{Fig_conf_1000_2000}, $N=5000$ in Figs.~\ref{Fig_conf_5000} and ~\ref{Fig_conf_5000_detail}, $N=10000$ in Fig.~\ref{Fig_conf_10000}, $N=14180$ in Fig.~\ref{Fig_conf_14180} and $N=40886$ in Fig.~\ref{Fig_conf_40886}~\footnote{The curiosity of the reader may be fulfilled by looking at the supplemental material of the present paper~\cite{AZ_suppA, AZ_suppB}.}.

There are interesting aspects emerging from these figures: although the internal topological charge is not appreciable until getting sufficiently close to the border (see Fig.~\ref{Fig_Qtopo}), in the central part of the disk we observe rather long "chains" of defects, typically sequences of heptagonal and pentagonal Voronoi cells, which contribute to lower the total energy. The length of these "chains" also is seen to increase with $N$. On the border of the disk, we typically see the presence of pentagonal and quadrilateral cells, almost perfectly alternating (for the largest configurations it is very difficult to observe due to the tiny size of these peripheral cells). As we mentioned earlier, this behavior is responsible of the scaling of $Q_{int} \approx 0.5 N_b$.

In particular we want to draw the attention of the reader to Fig.~\ref{Fig_conf_5000}, where we have plotted the two low energy configurations obtained with the D\&C algorithm for $N=5000$ (the one in the left plot is actually the lowest energy configuration that we have found). Both configurations have been generated in the same run of the algorithm and look the same at first sight. Indeed the structure of the defects that are well separated from the border is very similar. In Fig.~\ref{Fig_conf_5000_detail} we show a detail of the two figures, which clearly shows that the defect structure close to the border is slightly different. Small changes in the positions of the peripheral charges (not necessarily on the border) can lead to sensible changes in the local structure of the Voronoi cells, since in this region the inter-particle distance is quite small. These small changes may also produce much larger changes in the total energy of the system than changes performed in the central region. We regard this behavior as a further manifestation of the importance of the border in determining minimum energy configurations.

We also notice that all the Voronoi diagrams of these figures contain at most octagonal cells (this contrasts with the configurations in \cite{Olvera13} which seem to contain nonagons as well).  

Of course we don't believe that all our configurations for large $N$ are actual global minima  of the problem (the difficulty of finding a global minimum for such large configurations can only be underestimated) but we are confident that many features of the configurations we have found may resemble properties of the classical ground state.

\section{Conclusions}
\label{sec:concl}

We have studied the Thomson problem inside a disk and found a large number of configurations up to values of  $N$ (number of charges) which had never been considered before ($N \geq 10000$). To be able to work with such gigantic configurations we have devised a divide and conquer algorithm that proves particularly efficient for $N \geq 1000$ (as a matter of fact all our best configurations for $N \geq 1000$ have been found with D\&C).
For smaller $N$ we have applied a minimization algorithm that uses either the truncated Newton method (TN) or the basin--hopping method.
In all our numerical explorations we have enforced that the minimization be carried out with a {\sl fixed} number of border charges (as originally done in  \cite{Amore17}), which allows one to considerably speed up the numerical calculation.

Obtaining low energy configurations of large numbers of charges inside a domain is relevant for studying how the discrete system is approaching the continuum limit and possibly could allow a direct comparison with the models based on the continuum. In this respect we have observed that the charge density indeed tends to the expected continuum density for large enough $N$. Finally, we have observed that the Voronoi diagrams of the configurations
become increasingly complex as $N$ grows, with the appearance of large chains of alternating (pentagonal-heptagonal) cells: the topological charge of the domain, however, is mostly concentrated close to the border of the disk. We also found that the total internal topological charge does not scale as $N$, as reported in \cite{Olvera13}, but rather as $N_b$.

Although our results are specific to the disk, they can be generalized to study the Thomson problem in arbitrary domains of the plane (or even domains with curvature), with minor adaptations. We plan to conduct experiments on more general domains in future work.

\section*{Acknowledgements}
The authors acknowledge useful conversations with professors Isabel Dominguez Jimenez, Adil Mughal and David J. Wales. 
The research of P.A. was supported by Sistema Nacional de Investigadores (M\'exico). 
The plots in this paper have been plotted using Mathematica~\cite{wolfram} and {\rm MaTeX} \cite{szhorvat}. 
Numerical calculations have been carried out using python ~\cite{python} and numba ~\cite{numba}.

\end{document}